\documentclass[aps,twocolumn,showpacs,floatfix,amssymb]{revtex4}

\usepackage{graphicx}
\begin{document}

\title{Magnetic order and spin excitations  in  layered Heisenberg
antiferromagnets\\ with compass-model anisotropies}

\author{ A.A. Vladimirov$^{a}$, D. Ihle$^{b}$ and  N. M. Plakida$^{a}$ }
 \affiliation{ $^a$Joint Institute for Nuclear Research,
141980 Dubna, Russia}
 \affiliation{$^{b}$ Institut f\"ur Theoretische Physik,
 Universit\"at Leipzig,  D-04109, Leipzig, Germany }

\date{\today}

\begin{abstract}
The spin-wave excitation spectrum, the magnetization, and the N\'{e}el
temperature for the quasi-two-dimensional spin-1/2 antiferromagnetic Heisenberg model
with compass-model interaction in the plane proposed for iridates are calculated in the
random phase approximation. The spin-wave spectrum agrees well with data of Lanczos
diagonalization. We find that the N\'{e}el temperature is enhanced by the compass-model
interaction and is close to the experimental value for Ba$_2$IrO$_4$.
\end{abstract}


\maketitle

Spin-orbital physics in  transition-metal oxides has been
extensively studied  in recent years. A number of theoretical
models was proposed to describe a complicated nature of phase
transitions induced by competing  spin and  orbital interactions
as originally was considered in  Ref.~\cite{Kugel82}.  Whereas
the isotropic spin interaction can be treated within the
conventional  Heisenberg model, to study the orientation-dependent
orbital interaction the compass model is commonly used. The
latter reveals a large degeneracy of ground states resulting in a
complicated phase diagram.  In particular, quantum and
thermodynamic phase transitions in the two-dimensional (2D)
compass model were studied in
Refs.~\cite{Orus09,Wenzel08,Wenzel0}, where a first-order
transition was found for the symmetric compass model. A
generalized 2D Compass-Heisenberg (CH) model was introduced in
Ref.~\cite{Trousselet10}, where an important role of the spin
Heisenberg interaction in lifting the high degeneracy of the
ground state of the compass model was stressed. In
Ref.~\cite{Trousselet12} a  phase diagram of the CH model and
excitations within Lanczos exact diagonalization for finite
clusters on a square lattice were considered in detail. In
particular, spin-wave excitations and column-flip excitations in
nanoclusters characteristic to the compass model were analyzed.

A  strong relativistic spin-orbital coupling reveals  a
compass-model type interaction in 5$d$ transition metals. In
particular, it was shown in Ref.~\cite{Jackeli09}, that a strong
spin-orbit coupling in such compounds as Sr$_2$IrO$_4$ and
Ba$_2$IrO$_4$ results in  an effective antiferromagnetic (AF)
Heisenberg model for the pseudospins $1/2$  with the
compass-model anisotropy. The model can be used to explain the AF
long-range order (LRO)  below the N\'{e}el temperature $T_N =
230$~K in Sr$_2$IrO$_4$ and $T_N = 240$~K in Ba$_2$IrO$_4$ (see,
e.g.,~\cite{Boseggia13}). The spin-wave spectrum measured by
magnetic resonance inelastic x-ray scattering (RIXS) in
Sr$_2$IrO$_4$  shows a dispersion similar to that one in the
undoped cuprate La$_2$CuO$_4$~\cite{Kim12}.

In the present paper we calculate the spin-wave excitation
spectrum and magnetization for a  layered AF Heisenberg model with
anisotropic compass-model interaction in the plane. To take into
account the finite-temperature renormalization of the spectrum and
to calculate the N\'{e}el temperature $T_N$, we employ the
equation of motion method for the Green functions (GFs) for  spin
$S =1/2$ using the random phase approximation
(RPA)~\cite{Tyablikov75}. The results are compared with
experimental data for iridates and theoretical studies of the  2D
CH model in Ref.~\cite{Trousselet10} and in
Refs.~\cite{Igarashi13,Igarashi14}.

We consider the layered Heisenberg AF with the compass-model
interaction in the plane. The Hamiltonian  of the model  can be
written as
\begin{eqnarray}
  H & = & \frac{1}{2} \sum_{i, j}\Big\{  J_{ij}{\bf S}_{i}{\bf S}_{j} +
 \Gamma^x_{ij}  S^x_{i}{S}^x_{j} +
 \Gamma^y_{ij} {S}^y_{i}{S}^y_{j}\Big\}.
    \label{c1}
\end{eqnarray}
Here  $\,  J_{ij} =  J\,(\delta_{{\bf r}_j, {\bf r}_i \pm {\bf
a}_x} +
 \delta_{{\bf r}_j, {\bf r}_i \pm {\bf a}_y})
   + J_z\,\delta_{{\bf r}_j, {\bf r}_i \pm {\bf c}}$,
where  $J\,$ is the exchange interaction between the nearest neighbors (NN) in the plane
with the lattice constants $a_x =  \,a_y =a$, and $J_z$ is the coupling between the
planes with the distance $ c $. The compass model interaction is given by $\,
\Gamma^x_{ij} =
 \Gamma_x\,\delta_{{\bf r}_j, {\bf r}_i \pm {\bf a}_x}, \;
 \Gamma^y_{ij} = \Gamma_y \delta_{{\bf r}_j, {\bf r}_i \pm
{\bf a}_y} $.  The  {\it ab initio} many-body quantum chemistry
calculations give the following parameters for Ba$_2$IrO$_4$:
 $\,J = 65$~meV, $\,\Gamma_x = \Gamma_y = \Gamma =3.4$~meV, and
$J_z \gtrsim  3 - 5~\mu$eV~\cite{Katukuri14}. To compare our results with the theoretical
studies of the 2D CH model in Ref.~\cite{Trousselet10}, we consider also large
anisotropic compass-model interactions, $\, \Gamma_x > \Gamma_y > J$. In
Refs.~\cite{Igarashi13,Igarashi14} the spin-wave spectrum was calculated for a similar
model (\ref{c1}) in the linear spin-wave theory (LSWT), where in addition to the
isotropic exchange interaction $J_{ij}$ between the  NN in  (\ref{c1}) the next-nearest
neighbor (NNN) interaction  was also taken into account. We consider this more general
model in the Appendix.

We adopt a two-sublattice $(A,B)$ representation for the  AF LRO
below the N\'{e}el temperature. Then the Hamiltonian (\ref{c1})
with $\,\Gamma_x = \Gamma_y > 0\,$ is an easy-plane AF, where the
direction of the AF order parameter (OP) -- the magnetization of
one sublattice  in the $(x,y)$ plane -- is degenerate. To lift the
degeneracy, we assume anisotropic compass-model interactions
$\,\Gamma_x > \Gamma_y > 0 \,$. In this case the model (\ref{c1})
describes an easy-axis AF with the OP $\, \langle S_{i\subset A}^x
\rangle = - \langle S_{i\subset B}^x \rangle\, $ fixed along the
$x$ axis. We can consider also the limiting case, $\,\Gamma_x =
\Gamma_y \,$.  The AF LRO can be described by the AF wave vector
${\bf Q} = (\pi/a, \pi/a, \pi/c)$.

It is convenient to write the Hamiltonian (\ref{c1}) in terms of
the circular components $ {S_{i}^\pm} = {S_{i}^{y}} \pm  i
{S_{i}^{z}}$  in the form
\begin{eqnarray}
  {H}& = &\frac{1}{2}  \sum_{\langle i, j \rangle}
     \Big\{ J^x_{ij}\,{S^{x}_{i}}{ S^{x}_{j}}+ J^y_{ij}\frac{1}{2}\,
     \left[S^{+}_{i} S^{-}_{j}+ S^{-}_{i} S^{+}_{j} \right]
     \nonumber \\
&+&    \frac{1}{4} \Gamma^y_{ij}\,
     \left[S^{+}_{i} S^{+}_{j} + S^{-}_{i} S^{-}_{j}  \right] \Big\},
  \label{c2}
\end{eqnarray}
where $\,  J^x_{ij}= J_{ij} +  \Gamma^x_{ij}, \; J^y_{ij} =
J_{ij} + (1/2) \Gamma^y_{ij} \,$.

To calculate the spin-wave spectrum  of transverse spin
excitations, we introduce the retarded two-time commutator
GFs~\cite{Zubarev60}:
\begin{eqnarray}
 G_{nm}^{\alpha,\beta} (t-t') &= & -i \theta(t-t')
 \langle [ S_n^\alpha(t), S_m^\beta(t')]\rangle
 \nonumber \\
&=&
\int_{-\infty}^{+\infty}\,\frac{d\omega}{2\pi}\,e^{-i\omega(t-t')}
   \langle \! \langle  S_n^\alpha |
 S_m^\beta \rangle\! \rangle_\omega ,
     \label{r1}
\end{eqnarray}
where  $ \alpha,\beta = ( \pm )$, and $\,\langle \ldots
\rangle\,$ is the statistical average.  The indexes $n, m$ run
over $N/2$ lattice sites $i\, (j)$ in the sublattice $A\, (B)$.

There are four types of the GFs due to the two-sublattice
representation for normal and anomalous GFs which can be written
as  $4\times4$ matrix GF
\begin{eqnarray}
\hat G(\omega)= \langle\! \langle \left(
\begin{array}{c}
 S^+_i \\
 S^-_i \\
 S^-_j \\
 S^+_j
 \end{array}\right)  \mid \left(
S^-_{i'} \,S^+_{i'} \, S^+_{j'}\, S^-_{j'}\right)
 \rangle \! \rangle_{\omega}.
\label{r2}
\end{eqnarray}
Here the lattice sites $i, i'$ refer to the sublattice $A$ while
the lattice sites $j, j'$ refer to the sublattice $B$.

Using  equations of motion for spin operators, $\,
   i (d/dt) S^\pm_i(t) = [S^\pm_i , H ] =  \mp \sum_{n} \, J^{x}_{in}
  S_i^\pm S_n^x \pm \sum_{n} \,[ J_{in}^{y}\,S_i^x S_n^\pm  + (1/2)
\Gamma^{y}_{in}\,S_i^x S_n^\mp] \,$,  we obtain a system of
equations for the matrix components of the  GF~(\ref{r2}). In
particular,
\begin{eqnarray}
&&\omega \langle \! \langle  S_i^+ |
 S_{i'}^- \rangle\! \rangle_\omega =
 2\langle S_i^x \rangle\, \delta_{i,i'}
- \sum_{n} \, J^{x}_{in} \langle \! \langle \,
 S_i^+ S_n^x | S_{i'}^- \rangle\! \rangle_\omega
\nonumber  \\
 && +\sum_{n} [\,J_{in}^{y} \,
  \langle \!\langle \,
  S_i^x S_n^+ | S_{i'}^-  \rangle\! \rangle_\omega
  + (1/2)\Gamma^{y}_{in} \langle \!\langle \, S_i^x S_n^- | S_{i'}^-
\rangle\! \rangle_\omega ],
 \nonumber
\end{eqnarray}
\begin{eqnarray}
&&\omega \langle \! \langle  S_j^- |
 S_{j'}^+ \rangle\! \rangle_\omega =
- 2\langle S_j^x \rangle\, \delta_{j,j'} + \sum_{m} \, J^{x}_{jm}
\langle \! \langle \,
 S_j^- S_m^x |  S_{j'}^+  \rangle\! \rangle_\omega
\nonumber \\
 &&  - \sum_{m}[ \,J^{y}_{jm} \,
  \langle \!\langle \,
  S_j^x S_m^- |  S_{j'}^+  \rangle\! \rangle_\omega
   + \, (1/2)\Gamma^{y}_{jm} \langle \!\langle \, S_j^x S_m^+ |  S_{j'}^+
\rangle\! \rangle_\omega]. \nonumber
\end{eqnarray}
In the RPA~\cite{Tyablikov75}  for all GFs the following approximation  is used for the
lattice sites $n \neq i, \,m\neq j$, as e.g.,
\begin{eqnarray}
\langle \! \langle S_i^x S_n^\alpha | S_{i'}^\beta \rangle\!
\rangle_\omega
  & = & \langle S_i^x\rangle\, \langle \! \langle S_n^\alpha | S_{i'}^\beta  \rangle\! \rangle_\omega =
\sigma\, \langle \! \langle S_{n}^\alpha  | S_{i'}^\beta
\rangle\! \rangle_\omega, {}
\nonumber  \\
{} \langle \! \langle S_n^x S_i^\alpha | S_{i'}^\beta \rangle\!
\rangle_\omega & = & \langle S_n^x\rangle\, \langle \! \langle
S_i^\alpha | S_{i'}^\beta  \rangle\! \rangle_\omega = - \sigma\,
\langle \! \langle S_{i}^\alpha  | S_{i'}^\beta \rangle\!
\rangle_\omega ,\quad
 \label{r4}
\end{eqnarray}
where  $\langle S_i^x\rangle =  \sigma $ for $i \in A$ while
$\langle S_n^x\rangle = - \sigma $ for $n \in B$. A similar
approximation is used for the $B$ sublattice, where $\langle
S_j^x\rangle = - \sigma $ for $j \in B$ while $\langle
S_m^x\rangle =  \sigma $ for $m \in A$. The RPA results in a
closed system of equations for the components of the matrix GF
(\ref{r2}).

To solve the obtained system of equations  we introduce the
Fourier representation of spin operators for $N/2$ lattice sites
in two sublattices,   $\,  S_{i}^{\pm} = \sqrt{{2}/{N}}
  \sum_{\bf q}\,   S^{\pm}_{\bf q}\, \exp({\pm}i{\bf q} {\bf r}_i )\;$  and $\,
 S_{j}^{\pm} = \sqrt{{2}/{N}} \sum_{\bf q'}\,   S^{\pm}_{\bf q'}\,
 \exp({\pm}i{\bf q'} {\bf r}_j )\, $, where ${\bf q}$ and ${\bf q'}$  run over
$(N/2)$ wave vectors in the reduced BZ of each sublattice. Using
this transformation the equation  for the Fourier representation
of the matrix GF (\ref{r2})  can be written in the from
\begin{equation}
 \hat G({\bf q}, \omega) =   \{\omega \hat I -
  \hat V({\bf q})\}^{-1}\times  2\sigma\,\hat I_1 ,
\label{r5}
\end{equation}
where  $\hat I$ is the unity matrix,  $\hat I_1$ is a diagonal
matrix with the elements $\, d_{11} = d_{33} = 1$ and $\,d_{22} =
d_{44} = - 1$, and the interaction matrix  is given by
\begin{equation}
\hat V({\bf q}) =\left(
\begin{array}{cccc}
     A \qquad \;  0 \quad  B({\bf q})\quad C({\bf q})  \\
  \;0 \quad -A \; -C({\bf q})\;\, -B({\bf q})  \\
 \hspace{-6mm}  B({\bf q})\; \; C({\bf q}) \quad  A \; \qquad  \;0  \\
     \hspace{-7mm} -C({\bf q})\, - B({\bf q})\;\, 0 \quad - A  \\
         \end{array}\right).
\label{r6}
\end{equation}
Here the interaction parameters are:
\begin{eqnarray}
 A & = & \sigma\, J^x(0)= \sigma\,[J(0) + 2\, \Gamma_x] ,
\nonumber \\
J({\bf q}) & = & 2 J \,(\cos{ q}_x + \cos{ q}_y) + 2 J_z\,\cos{
q}_z,
\nonumber \\
B({\bf q}) & = & \sigma\, \Gamma_y \cos{ q}_y, \quad
  C({\bf q}) = \sigma\,[ J({\bf q})+ \Gamma_y \cos{ q}_y ].
    \label{r7}
\end{eqnarray}
The spectrum of spin waves is determined from the equation
\begin{equation}
 {\rm Det}\, |\omega \hat I -
  \hat V({\bf q})| = 0  .
 \label{r8}
 \end{equation}
After some algebra we obtain the biquadratic equation for the
frequency  $\omega$ of spin-wave excitations:
\begin{eqnarray}
&&\omega^4 - 2 \omega^2[A^2 +B^2({\bf q}) -C^2({\bf q})]+[B^2({\bf
q}) - C^2({\bf q})]^2
\nonumber \\
&& - 2 A^2 [C^2({\bf q}) + B^2({\bf q})]  + A^4 = 0 .
 \nonumber
\end{eqnarray}
The solution of this equation reads
\begin{eqnarray}
 \omega_{\nu}({\bf q})
& = &\pm \{A^2 +B^2({\bf q}) -C^2({\bf q}) + 2 \nu A\,B({\bf
 q})\}^{1/2}
 \nonumber \\
 &\equiv &\pm \sigma\,\varepsilon_\nu({\bf q}),
\label{r10}
 \end{eqnarray}
where $\nu = \pm 1$. The energy of excitations  for ``acoustic'' $\,\varepsilon_{-}({\bf
q})\,$ and ``optic'' $\,\varepsilon_{+}({\bf q})\,$ modes  are
\begin{eqnarray}
 \varepsilon_{-}({\bf q}) & = & \Big\{ J^2(0)
  - J^2({\bf q}) + 4\,\Gamma_x [J(0)\,  + \Gamma_x]
-  \nonumber \\
  &- & 2\,\Gamma_y\, [J(0) +J({\bf q})
 +  2\, \Gamma_x] \, \cos{q}_y \Big\}^{1/2} ,
 \label{r11a} \\
 \varepsilon_{+}({\bf q}) & = & \Big\{ J^2(0)
  - J^2({\bf q}) + 4\,\Gamma_x [J(0)  +  \Gamma_x] +
\nonumber\\
 & + & 2\,\Gamma_y [ J(0)- J({\bf q})
 +  2\, \Gamma_x] \, \cos{q}_y \Big\}^{1/2}.
\label{r11b}
\end{eqnarray}
These two branches   are coupled by the relation $\,
\varepsilon_{-}({\bf q + Q})= \varepsilon_{+}({\bf q})
 \,$ for the AF  wave vector ${\bf Q}$.

For the symmetric compass-model interaction, $\, \Gamma_x =
\Gamma_y = \Gamma$, for ${\bf q} = 0$ we have the gapless
acoustic mode while the optic mode has a gap:
\begin{eqnarray}
 \varepsilon_{-}(0) &= & 0, \quad
 \varepsilon_{+}(0) = 2 \sqrt{ \,\Gamma\, J(0)
  + 2\Gamma^2 }  > 0  .
\label{r12b}
\end{eqnarray}
For  the wave vector ${\bf q} = {\bf Q}$ we have the opposite
results: $\, \varepsilon_{-}({\bf Q}) = \varepsilon_{+}(0) > 0,
\quad \varepsilon_{+}({\bf Q}) = \varepsilon_{-}(0) = 0 \,$.  In
the anisotropic case $\Gamma_x
> \Gamma_y $ the spectrum of excitations  has  gaps both at ${\bf q} = 0$
and ${\bf q} = {\bf Q}$:
\begin{equation}
 \varepsilon_{-}(0) = \varepsilon_{+}({\bf Q}) =
 2\sqrt{(\Gamma_x - \Gamma_y)\, (J(0)\,  + \Gamma_x)} .
 \label{r13}
\end{equation}
For a conventional AF Heisenberg model with $\,\Gamma_x =
\Gamma_y =0 \,$ we have only one branch with the  dispersion
$\varepsilon_{-}({\bf q}) = \varepsilon_{+}({\bf q}) =
\sqrt{J^2(0) - J^2({\bf q})}$ which is gapless both at ${\bf q} =
0$ and ${\bf q} = {\bf Q}$.

A similar equation of motion method for the matrix GF (\ref{r2}) can be employed in the
LSWT  using the transformation $\,S_{i}^+ = \sqrt{2S}\, a_i, \; S_{i}^- = \sqrt{2S}\,
a_i^\dag, \; S_{i}^x = S -  a_i^\dag a_i\,$ for the sublattice $A$ and the similar
transformation for the sublattice $B$ ($a_i \rightarrow  b_i^\dag$). Then keeping only
linear terms in the bose-like operators $(\,a_i,\, a_i^\dag)$ and $ (\,b_i,\, b_i^\dag) $
we obtain Eqs.~(\ref{r10}), (\ref{r11a}), (\ref{r11b})  for the spin-wave spectrum in
LSWT  with the sublattice magnetization $\sigma $ substituted by spin $S$. The same
spectrum in LSWT was obtained in Refs.~\cite{Trousselet10,Trousselet12}. Note that in the
RPA the energy of spin excitations $\omega_{\pm}({\bf q})$, Eq.~(\ref{r10}), is reduced
in comparison with the LSWT since $\, \sigma < S \, $ even at zero temperature due to
zero-point fluctuations in the AF state. The spectrum (\ref{r10}) for the symmetric
compass model, $\,\Gamma_x =\Gamma_y \,$, is similar to the spectrum of the anisotropic
AF Heisenberg model considered in Ref.~\cite{Rudoy74} and
Refs.~\cite{Igarashi13,Igarashi14}.

\begin{figure}
\resizebox{0.35\textwidth}{!}{%
\includegraphics{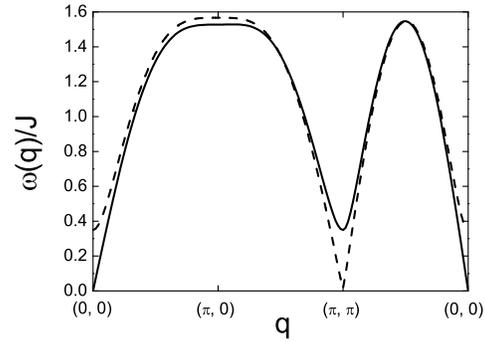}}
\caption{ Spectrum of spin-wave excitations $\omega_{-}({\bf q})$ (bold line) and $
\omega_{+}({\bf q})$ (dashed line)  along the symmetry directions in the BZ for the
symmetric compass model with $\,\Gamma_x =\Gamma_y = \Gamma = 0.052 \, J$ and $\,J_z
=0$.}
 \label{fig1}
\end{figure}

In Figure~\ref{fig1} the spectrum of spin waves $\, \omega_{\pm}({\bf q})\,$ in the plane
in RPA for the parameters $J = 65$~meV, $\Gamma =3.4$~meV found for
Ba$_2$IrO$_4$~\cite{Katukuri14} is shown  at $T = 0$. The spectrum $\omega_{-}({\bf q})$
shows a gap at the wave vector ${\bf Q}$ given by $\, \omega_{-}({\bf Q})=
2\,\sigma\,\sqrt{\Gamma\, J(0) + 2\Gamma^2} \approx 1.48 \, J \sqrt{ \Gamma/ J } \approx
22$~meV for  $\sigma = 0.37$. This value is comparable with the maximum energy of
excitations $\omega_{-}^{\rm max}({\bf Q}/2) = 4\,\sigma \,J \sqrt{1 + \Gamma/J }\approx
1.5\, J$ that gives $\omega_{-}({\bf Q})/\omega_{-}^{\rm max}({\bf Q}/2) \approx 0.22$.
We can suggest that the spin-wave spectrum in Ba$_2$IrO$_4$ should be similar to that one
measured by RIXS in Sr$_2$IrO$_4$~\cite{Kim12}.  The latter was fitted by a one-branch
phenomenological $\, J -J' - J''\,$ model with $\, J = 60$~meV, $\, J' = - 20$~meV,  and
$\, J'' = 15$~meV. The spectrum does not reveal a gap  in the acoustic branch
$\omega_{-}({\bf q})$ at ${\bf Q}$ as for Ba$_2$IrO$_4$. However, since the intensity of
scattering on magnons is proportional to $\,1/ \omega({\bf q})\,$, strong scattering on
the gapless branch  $\, \omega_{+}({\bf q}) \rightarrow 0 \,$ for ${\bf q} \rightarrow
{\bf Q}$ completely suppresses scattering on the gapped $\, \omega_{-}({\bf q})$ branch.
To distinguish scattering on the two branches, high-resolution studies are necessary. A
possibility of observation of a two-peak structure in the  RIXS experiments caused by the
two-branch spectrum of spin waves  is discussed in
Refs.~\cite{Igarashi14,Igarashi14a,Igarashi14b}.  For the model (\ref{c1}) with the
exchange interaction $J_{ij}$ only between nearest neighbors the energy of excitations at
$\,{\bf q}_1 = (\pi/2, \pi/2)\,$, $\, \omega_{-}({\bf q}_1) = \omega_{+}({\bf q}_1)$, is
nearly equal to $\, \omega_{\pm}({\bf q} = \pi, 0)$ (up to $\,\pm \Gamma / J\,$), while
in the RIXS experiment $\, \omega({\bf q}_1) \approx (1/2)\omega({\bf q} = \pi, 0)$ was
found. By taking into account the exchange interactions also between the NNN in the plane
the spectrum can be fitted to the experimentally observed one  as discussed in the
Appendix.

\begin{figure}
\resizebox{0.35\textwidth}{!}{%
\includegraphics{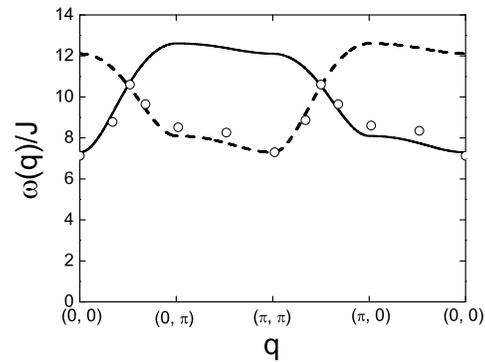}}
\caption{Spectrum of spin-wave excitations $\omega_{-}({\bf q})$
(bold line) and $ \omega_{+}({\bf q})$ (dashed line)   along the
symmetry directions in the BZ for the anisotropic  compass model
with $\,\Gamma_x = 8.9\, J, \quad \Gamma_y = 4.5 \, J$, $\,J_z
=0$. Circles  are numerical results from
Ref.~\cite{Trousselet10}. }
 \label{fig2}
\end{figure}

Figure~\ref{fig2} shows the spin-wave dispersion for large
anisotropic interaction, $\,\Gamma_x = 8.9\, J, \quad \Gamma_y =
4.5 \, J$ used in Ref.~\cite{Trousselet10} in numerical
calculations  with Lanczos exact diagonalization. Our RPA
calculations give a similar  formula for the spectrum as in LSWT
except for the prefactor $\sigma = 0.44$ instead of $S = 1/2$ in
LSWT. The dispersion curves  are in good agreement with numerical
ones shown by circles which were multiplied by the factor 10/4,
since in Ref.~\cite{Trousselet10}, instead of spin $1/2$
operators, the Pauli matrices are used so that the exchange
integral $I$ corresponds to our $ (1/4)\, J$ in Eq.~(\ref{c1}),
and in Fig.~(4) of Ref.~\cite{Trousselet10} the energy unit is
$J_c = 10 I $. The spectrum reveals a large gap at all wave
vectors caused by the large value of $\,\Gamma_x\,$ and a
noticeable dispersion only along the $\Gamma (0,0) \rightarrow
Y(0, \pi)$ direction due to a large, in comparison with $J$,
interaction $\,\Gamma_y = 4.5\, J$.

To calculate the sublattice magnetization $\sigma = \langle S_i^x
\rangle $ in RPA, we use the kinematic relation $\, S_i^x = (1/2)
- S_i^- S_i^+\,$ for spin $\,S=1/2$ which results in  the
self-consistent equation
\begin{equation}
 \sigma =  \frac{1}{2}-\frac{1}{N/2}
  \sum_{\bf q}\,\langle S_{\bf q}^-  S_{\bf q}^+ \rangle .
 \label{m1}
\end{equation}
The  spin correlation function $\,  \langle S_{\bf q}^-  S_{\bf
q}^+ \rangle $ can be calculated from the  GF $\, \langle \!
\langle S_{\bf q}^+ | S_{\bf q}^- \rangle\! \rangle_\omega\,$
which  follows from the GF (\ref{r5}):
\begin{eqnarray}
\langle \! \langle S_{\bf q}^+ | S_{\bf q}^- \rangle\!
\rangle_\omega & = & 2\sigma\,\frac{a_{\bf q}(\omega)}{[\omega^2
- \omega_{-}^2({\bf q})]
  [\omega^2 - \omega_{+}^2({\bf q})]},
 \label{m2}\\
 a_{\bf q}(\omega) &= &\omega^3 + A \, \omega^2
 - [A^2 + B^2({\bf q})- C^2({\bf q})]\, \omega -
\nonumber \\
& - & A^3 + A\,[B^2({\bf q})+ C^2({\bf q})].
\nonumber
\end{eqnarray}
Using the spectral representation for GFs, for  the correlation
function   we obtain
\begin{equation}
\langle S^-_{\bf q}  S^+_{\bf q}\rangle = 2\sigma\,\sum_{\mu,\nu =
\pm 1} I_{\mu\nu}({\bf q}) \, N(\mu \omega_{\nu} ({\bf q}))\,,
 \label{m3}
\end{equation}
where $N(\omega)= [\exp (\omega/T) - 1]^{-1}$, and the
contribution from the four poles of the GF (\ref{m2}) is given by
\begin{equation}
I_{\mu\nu}({\bf q}) = \frac{a_{\bf q}( \mu \omega_{\nu} ({\bf
q}))}{8 \mu \nu \omega_{\nu} ({\bf q})\, AB({\bf q})} .
 \label{m4}
\end{equation}
Note that $I_{\mu\nu}({\bf q})$ does not depend on $\sigma$.

Using relation (\ref{m3}) we perform the self-consistent solution
of Eq.~(\ref{m1}) for the magnetization $\, \sigma \,$.
\begin{figure}[ht!]
\resizebox{0.35\textwidth}{!}{%
\includegraphics{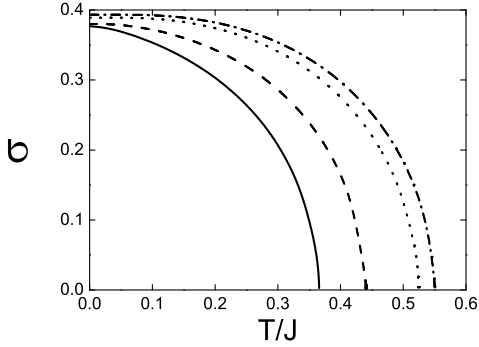}}
\caption{ Sublattice magnetization $\sigma = \langle S_i^x
\rangle $ for the parameters$\, J_z = 5\cdot 10^{-5}\, J,
\,\Gamma_x =  0.052 \,J\,$ for $\Gamma_y/\Gamma_x  = 1\,$ (solid
line),  0.95 (dashed line), 0.5 (dotted),  and for
$\Gamma_y/\Gamma_x \leqslant 0.1\,$ (dash-dotted).}
 \label{fig3}
\end{figure}
Figure~\ref{fig3} shows the sublattice magnetization for  $ \,
J_z = 5\cdot 10^{-5}\, J, \,\Gamma_x =  0.052 \,J$ for various
$\Gamma_y/\Gamma_x$.  For the symmetric compass model, $\Gamma_x =
\Gamma_y = 0.052\,J$, the N\'{e}el temperature $T_N = 0.365 J =
275$~K  is close to $T_N = 240$~K observed in experiment for
Ba$_2$IrO$_4$. We stress that the anisotropy of the compass-model
interaction, $\Gamma_y/\Gamma_x < 1 \,$, enhances $T_N$.

To study  the $T_N$ dependence on the parameters of the model,  we
consider Eq.~(\ref{m1}) in the limit $\sigma \rightarrow 0$. In
this limit $N(\omega_{\nu}) \approx (T/\sigma \varepsilon_{\nu})$,
and  for the N\'{e}el temperature  we have the equation:
\begin{equation}
  \frac{1}{2} = \frac{1}{N/2}
 \sum_{\bf q}\, \sum_{\mu,\nu = \pm 1} I_{\mu\nu}({\bf
q}) \, \frac{2 \,T_N}{\mu \varepsilon_{\nu} ({\bf q})} .
 \label{m6}
\end{equation}
Therefore,
\begin{equation}
T_N = \frac{1}{4C},  \quad C = \frac{1}{N/2}
 \sum_{\bf q}\,\sum_{\mu, \nu}  \,
 \frac{I_{\mu\nu}({\bf q})}{\mu \varepsilon_{\nu}({\bf
q})}  .
 \label{m7}
\end{equation}
Let us study in which cases  the integral over ${\bf q}$ in
(\ref{m7}) has a finite value that results in a finite $T_N$.

At first we consider  the symmetric compass model, $\Gamma_x =
\Gamma_y = \Gamma $. In this case $\varepsilon_{-}({\bf q}) = 0$
at ${\bf q} = 0$ and $\varepsilon_{+}({\bf q}) = 0$ at ${\bf q} =
{\bf Q}$. Since these two branches are symmetric,  we can consider
only the divergency of the integral in (\ref{m7})  at ${\bf q} =
0$ for $\varepsilon_{-}({\bf q})$ given around ${\bf q} = 0$ by
\begin{eqnarray}
\varepsilon_{-}^2({\bf q})& = & 2(J(0) + \Gamma)\{J \, q_x^2
\nonumber \\
&+&  [ J + \Gamma^2/(J(0)+ \Gamma)\,]\, q_y^2 + J_z \,q_z^2 \}.
 \label{m8}
\end{eqnarray}
The integral in (\ref{m7}) diverges as $\int d^3{\bf q}
/{\varepsilon_{-}^2({\bf q})}$ if any coefficient before $q_x$,
$q_y$ or $q_z$ in (\ref{m8}) is zero. In particular, for nonzero
$J(0)$ there is no LRO at finite $T$ for $J_z = 0$.

In the limiting case $\Gamma \rightarrow 0$ we have $\, \lim
I_{\mu \nu}({\bf q}) =
 ({A + \mu \omega_{\bf q}})/( {4 \mu \omega_{\bf q}})\,$
with $ \omega_{\bf q} = \sqrt{A^2 - C^2({\bf q})} $. From
Eq.~(\ref{m7}) we get the conventional formula for $T_N$ of the AF
Heisenberg model (c.f. Ref.~\cite{Vladimirov13}) :
\begin{eqnarray}
 T_N(\Gamma=0) = \left\{\frac{8J(0)}{N}\sum_{\bf q}\,
 \frac{1}{J(0)^2 - J^2({\bf q})  }\right\}^{-1}.
 \label{c21}
 \end{eqnarray}
Thus, for a symmetric 2D compass model  we have no LRO at finite
$T$. To obtain LRO,  we must have finite values of both $J\,$ and
$ J_z\,$.
\begin{figure}
\resizebox{0.35\textwidth}{!}{%
\includegraphics{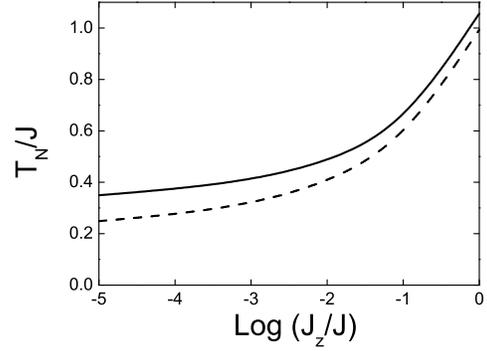}}
\caption{N\'{e}el temperature  $T_N$ as a function of $\,J_z $
with $\,\Gamma_x =\Gamma_y = 0.052 \, J$ (solid line) and
$\,\Gamma_x =\Gamma_y  = 0 $ (dashed line).}
 \label{fig4}
\end{figure}
\begin{figure}
\resizebox{0.35\textwidth}{!}{%
\includegraphics{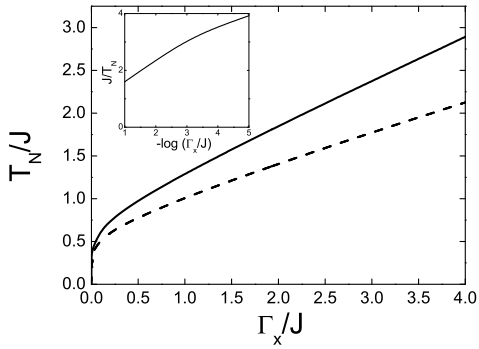}}
\caption{ N\'{e}el temperature  $T_N$ as a function of
$\,\Gamma_x$ for  $\, J_z = 0$, $\,\Gamma_y = 0.1\Gamma_x $ (solid
line) and  $\,\Gamma_y = 0.9\Gamma_x $ (dashed line). In the
inset the  $1/ T_N$ dependence  is shown in the logarithmic scale
for small $\,\Gamma_x$.}
 \label{fig5}
\end{figure}
The N\'{e}el temperature $T_N$ as a function of the interplane
coupling $J_z$ is shown in Fig.~\ref{fig4}  for the interaction
$\,\Gamma_x =\Gamma_y  = 0.052\,J\, $ and for $\,\Gamma_x
=\Gamma_y =0 $.  We can conclude that the compass-model
interaction enhances the N\'{e}el temperature and, in particular,
the anisotropy of the compass-model interaction results in a
further increase of $T_N$ as shown in Fig.~\ref{fig3}. In the
anisotropic case $\Gamma_x > \Gamma_y $ the spectrum of
excitations  has a gap at $q =0$, Eq.~(\ref{r13}), and therefore
neither branch of this spectrum ever reaches zero, so that we
have a finite $T_N$ even for  $J_z = 0$. Figure~\ref{fig5}
demonstrates  the dependence of $T_N$ on $\,\Gamma_x \,$ for $J_z
= 0$, $\,\Gamma_y = 0.1\,\Gamma_x $ and $\,\Gamma_y =
0.9\,\Gamma_x $. For $\,\Gamma_x \rightarrow 0\,$ the N\'{e}el
temperature goes to zero  as shown in the inset.\\

To summarize, we have studied the spin-wave spectrum for the
Heisenberg model with anisotropic compass-model interaction
within the RPA. The spectrum has  gaps at ${\bf q} = 0$ or at the
AF wave vector ${\bf Q}$ for nonzero compass-model interactions.
The calculation of the N\'{e}el temperature $T_N$ shows that for
the symmetric compass-model interaction, $\Gamma_x = \Gamma_y $,
and a nonzero exchange interaction $J$, the AF LRO at finite $T$
can exist only for a finite coupling  $J_z$ between the planes.
For the anisotropic compass-model interaction, $\Gamma_x >
\Gamma_y $, and a finite exchange interaction $J$ in the plane,
the AF LRO with finite N\'{e}el temperature  emerges even in the
2D case as observed in finite cluster
calculations~\cite{Trousselet10, Trousselet12}. In any case,
$T_N$ is enhanced by the compass-model interaction. \\

The authors would like to thank G. Jackeli,  A.\,M.~Ole\'{s}, J.~Richter, Yu.\,G. Rudoy,
and V.~Yushankhai  for valuable discussions. We thank T.\, Nagao who drews our attention
to Refs.~\cite{Igarashi13,Igarashi14,Igarashi14a,Igarashi14b}  for comments on the first
version of our paper ~\cite{Vladimirov14}. A financial support by the Heisenberg--Landau
Program of JINR is acknowledged.

\appendix

\section{Further distant  neighbors}

To fit  the spin-excitation spectrum observed by RIXS in Sr$_2$IrO$_4$  in
Ref.~\cite{Kim12} we consider a more general exchange interaction which includes the NNN
interaction  $\, J^{nn}_{ij} = J^{\prime}_{ij} + J^{\prime \prime}_{ij}$, where
$J^{\prime}_{ij} = J^{\prime} [\delta_{{\bf r}_j, {\bf r}_i \pm ({\bf a}_x+{\bf a}_y)}  +
\delta_{{\bf r}_j, {\bf r}_i \pm ({\bf a}_x-{\bf a}_y)}]\,$ and $\,
 J^{\prime \prime}_{ij} =  J^{\prime \prime}(\delta_{{\bf r}_j, {\bf r}_i \pm 2{\bf a}_x}
+  \delta_{{\bf r}_j, {\bf r}_i \pm 2{\bf a}_y})$.  Note that  in
the two-sublattice model for the Hamiltonian  (\ref{c1}) the
in-plane exchange interaction $J_{ij}$ acts between the NN  on
the two sublattices,  while for the NNN exchange interaction $
J^{nn}_{ij}$  the lattice sites $i $ and $j$ refer to the same
sublattice. Therefore, in the equation for the GF  (\ref{r5})
the  exchange interaction $ J^{nn}_{ij}$ gives a contribution only
for the diagonal terms in the interaction matrix (\ref{r6}). This
results in the transformation of the  parameter $A = \sigma
\,J^x(0)$ to the function $ A({\bf q}) = \sigma\, [ J^x(0)-
J^{nn}(0)+ J^{nn}({\bf q})]$, where  $\, J^{nn}({\bf q}) = 4
J^{\prime}\, \cos {q}_x \cos{q}_y   + 2 J^{\prime \prime} (\cos
2{q}_x + \cos 2{q}_y )$. The solution of Eq.~(\ref{r8}) yelds the
same spectrum of spin excitations~(\ref{r10}) with the parameter
$A$ substituted by $ A({\bf q})$.
 The energy of excitations  for ``acoustic'' $\,\varepsilon_{-}({\bf
q})\,$ and ``optic'' $\,\varepsilon_{+}({\bf q})\,$ modes  is
given by
 Eqs.~(\ref{r11a}) and (\ref{r11b}), where instead of $ J(0)$ we have to use the function
$ [J(0)- J^{nn}(0)+ J^{nn}({\bf q})]$. If we take $\Gamma_x = \Gamma_y = 0$, the spectrum
of spin waves transforms to $\varepsilon({\bf q}) =  \{[J(0)
 - J^{nn}(0) + J^{nn}({\bf q})]^2 - J^2({\bf q}) \}^{1/2}\,$ which  is gapless
both at ${\bf q}= 0$ and ${\bf q}= (\pi,\pi)$.
\begin{figure}[h!]
\resizebox{0.35\textwidth}{!}{%
\includegraphics{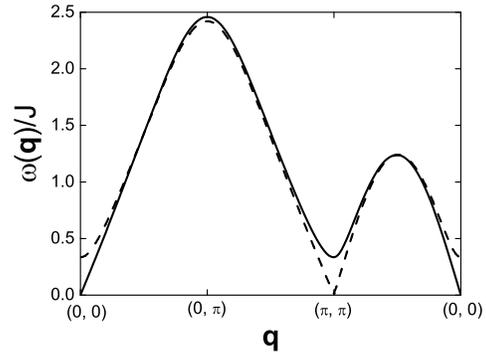}}
\caption{ Spectrum of spin-wave excitations $\omega_{-}({\bf q})$ (bold line) and $
\omega_{+}({\bf q})$ (dashed line)  along the symmetry directions in the BZ for the model
with $\,\Gamma_x =\Gamma_y = \Gamma = 0.052 \, J$, $\,J_z =0$, and NNN interactions
$J^{\prime} = - (1/3) J$, $J^{\prime \prime} = (1/4) J$.}
 \label{fig6}
\end{figure}

Taking into account the $ J^{nn}({\bf q})$ term with $J^{\prime}
 = - (1/3) J$ and  $J^{\prime \prime} = (1/4) J$ as suggested in experiment
~\cite{Kim12} and in Refs.~\cite{Igarashi13,Igarashi14} we obtain   the spin-wave
spectrum in RPA  shown in Fig.~\ref{fig6}. In comparison with Fig.~\ref{fig1} now the
excitation energy  $\, \omega(\pi/2, \pi/2)\approx (1/2)\omega(\pi, 0)$,  as observed in
the RIXS experiment and in the LSWT in Refs.~\cite{Igarashi13,Igarashi14}. The maximum
energy of excitation $\omega(\pi,0) = 2.5 J$ is larger than in Fig.~\ref{fig1}, where
$\omega(\pi,0) \approx 1.6 J $,  but it is still smaller than  the experimental value of
200 meV~\cite{Kim12} due to the renormalization of the spin-excitation energy in  RPA
given by  the reduced magnetization $\sigma = 0.36$   in comparison with $\sigma = S =
1/2$ in the LSWT. Large values of $J^{\prime}  = - (1/3) J$ and $J^{\prime \prime} =
(1/4) J$ in comparison with $J^{\prime} \approx - 0.1 J$ found in
La$_2$CuO$_4$~\cite{Coldea01} may be explained by a mixing of the $j_{eff} = 1/2 $ states
with higher energy  $j_{eff} = 3/2 $ states as suggested in
Refs.~\cite{Igarashi14a,Igarashi14b}, where an itinerant-electron multi-orbital model was
considered.

The NNN interaction also results  in the lowering of the N\'{e}el temperature, $T_N = 0.3
J = 220$~K (for $J = 65$~meV), in comparison with $T_N = 0.365 J = 275$~K found for  $\,
J^{nn}_{ij}= 0$  and is close to $T_N = 240$~K observed in experiment for Ba$_2$IrO$_4$.

\end{document}